\documentclass[twocolumn,prl,citeautoscript,amsmath,amssymb]{revtex4-1}
\usepackage{graphicx}
\usepackage{color}
\usepackage{upgreek}
\usepackage{float}
\usepackage{booktabs}
\usepackage{mathrsfs}
\usepackage{fancyhdr}
\usepackage{subcaption}
\usepackage{enumitem}
\usepackage[table,xcdraw]{xcolor}
\usepackage[breaklinks,colorlinks,citecolor=blue,linkcolor=blue]{hyperref}
\newcommand{\frob}[1]{\ensuremath{\left|\right| #1 \left|\right|_F}}

\newcommand{\add}[1]{{\color{black}#1}}

\AtBeginDocument{
\heavyrulewidth=.08em
\lightrulewidth=.05em
\cmidrulewidth=.03em
\belowrulesep=.65ex
\belowbottomsep=0pt
\aboverulesep=.4ex
\abovetopsep=0pt
\cmidrulesep=\doublerulesep
\cmidrulekern=.5em
\defaultaddspace=.5em
}
\begin{document}
\title{Performance optimization for drift-robust fidelity improvement of two-qubit gates}
\author{G. A. L. White$^\dagger$}
%\email{white.g@unimelb.edu.au}
\affiliation{$^\dagger$School of Physics, University of Melbourne, Parkville, VIC 3010, Australia\\
$^\ddag$ School of Mathematics and Statistics, University of Melbourne, Parkville, VIC, 3010, Australia}
\author{C. D. Hill$^{\dagger\ddag}$}
\affiliation{$^\dagger$School of Physics, University of Melbourne, Parkville, VIC 3010, Australia\\
$^\ddag$ School of Mathematics and Statistics, University of Melbourne, Parkville, VIC, 3010, Australia}
%\affiliation{$^\ddag$ School of Mathematics and Statistics, University of Melbourne, Parkville, VIC, 3010, Australia}
\author{L. C. L. Hollenberg$^\dagger$}
\email{lloydch@unimelb.edu.au}
\affiliation{$^\dagger$School of Physics, University of Melbourne, Parkville, VIC 3010, Australia\\
$^\ddag$ School of Mathematics and Statistics, University of Melbourne, Parkville, VIC, 3010, Australia}
\begin{abstract}
Quantum system characterization techniques represent the front line in the identification and mitigation of noise in quantum computing, but can be expensive in terms of quantum resources and time to repeatedly employ. Another challenging aspect is that parameters governing the performance of various operations tend to drift over time, and monitoring these is hence a difficult task. One of the most promising characterization techniques, gate set tomography (GST), provides a self-consistent estimate of the completely positive, trace-preserving (CPTP) maps for a complete set of gates, as well as preparation and measurement operators. We develop a method for performance optimization seeded by tomography (POST), which couples the power of GST with a classical optimization routine to achieve a consistent gate improvement in just a short number of steps within a given calibration cycle. By construction, the POST procedure finds the best available gate operation given the hardware, and is therefore robust to the effects of drift. Further, in comparison to other quantum error mitigation techniques, it builds upon a one-time application of GST. To demonstrate the performance of this method on a real quantum computer, we map out the operations of six qubit pairs on the superconducting \emph{ibmq\_poughkeepsie} quantum device. Under the restriction of logical-only control, we monitor the performance of the POST approach on a chosen CNOT gate over a period of six weeks. In this time, we achieve a consistent improvement in gate fidelity, averaging a fidelity increase of 21.1\% as measured by randomized benchmarking. The POST approach should find wide applicability as it is hardware agnostic, and can be applied at the upper logical level or at a deeper pulse control level. 
\end{abstract}
 
\maketitle
\section{Introduction}
The nascent field of quantum computing has seen an emergence of many experimentally realized small-scale devices in recent years, most notably in superconducting qubit systems and trapped atomic spins \cite{sandia-GST,Arute2019,IBM-53,intel-49,rigetti-19}. Different architectures have achieved high fidelity one and two-qubit gates, as well as the construction of multi-qubit entangled states \cite{Barends2014,Zeuner2018,PhysRevLett.117.140501,He2019,Hong2019, Kjaergaard2019,Bradley2019,Arute2019}. Despite this progress, current hardware cannot yet demonstrate large-scale topological quantum error correction below threshold, and there are many significant obstacles to overcome before qubit numbers can be scaled up to useful levels. Quantum computers presently face the challenge of imperfections in state preparation, measurement errors, and erroneous logical gates. Before improvement can be achieved, comprehensive characterization techniques are essential in mapping where deficiencies lie. \par 
Noise on real quantum devices is challenging to understand quantitatively. In particular, it is difficult to isolate device behaviour given the tendency of noise and system parameters to \emph{drift} \cite{Klimov2018,Fogarty2015,Chow2009}. This is one of the many barriers facing the improvement of quantum hardware. Common characterization techniques such as quantum process tomography (QPT) \cite{PhysRevLett.78.390} and randomized benchmarking (RB) \cite{PhysRevA.77.012307} offer an insight into the quality of a qubit, but suffer from respective self-consistency and limited information issues. Gate set tomography (GST), introduced in \cite{gst-2013,PhysRevA.87.062119}, provides a relatively novel method in which the preparation, gate, and measurement operations can be implemented in conjunction with each other and separately characterized. The results can be highly accurate, but with the trade-off that a large number of experiments are required to provide the data. The analysis itself is also computationally demanding. As a consequence, there are relatively few examples of two-qubit GST carried out experimentally in the literature \cite{osti_1428158,GST-improvement-china,Song2019}. \add{Importantly, GST puts a lens on otherwise inaccessible quantities such as the average gate infidelity, dominant noise channels, and its diamond norm. Here, we demonstrate how the comprehensive information provided by GST can be leveraged to reduce the coherent noise in quantum operations. We consider both the diamond norm of the gate, which may be extracted from the initial estimate, and the randomized benchmarking infidelity, which may be efficiently estimated in the tune-up procedure.} \par 
Two-qubit gates are the most significant source of error in many quantum circuits, and so minimising their infidelity is critical to the performance of quantum algorithms. In this manuscript, we develop a method for performance optimization seeded by tomography (POST) to consistently improve two-qubit CNOT gates based on a hybrid quantum-classical approach. 
We characterize the bare two-qubit logic gate using GST, find the optimal corrective parameters \add{for} bookend single-qubit unitaries, and then use these as a seed to the Nelder-Mead algorithm in order to find the best improvement for a given calibration cycle.
\add{We consider two regimes of control: corrective gates acting solely on the control qubit, and corrective gates acting both on the control and target qubits.}
Following the one-time overhead of GST, each daily optimization is performed in a small ($<$150) number of \add{single-length RB} experiments to overcome any drift which has occurred. 
\add{This is similar to the use of RB as an objective function in \cite{google-RB}, but with the incorporation of the more detailed information provided by GST. We discuss later the practical advantages to seeding this procedure with the initial GST estimates.}
We test the method on the \emph{ibmq\_poughkeepsie} quantum device\add{, a 20 qubit transmon device with a quantum volume of 16, CNOT error rates typically ranging between $1$ and $5$\%, and calibrations usually performed once per day. In particular, we measure the CNOT gate under consideration to have an RB infidelity $r_{\text{CNOT}}$ of 0.0337, diamond norm $\mathcal{E}_{\Diamond\text{CNOT}}$ of 0.0683, and unitarity (as defined in \cite{noise-coherence-2015}) of 0.955. The computational costliness of computing confidence factories on two-qubit GST means that we do not have error bars on these values. We discuss in the main body the reliability of the GST estimates, and how the procedure is relatively insensitive to these.} The coupling of the GST seed with classical optimization is successful at improving the gate. When tested on an experimental device, we find the POST approach is effective even weeks after the initial characterization. 
\add{At approximately 0.5 ms per shot, this procedure requires about 5 seconds per shot of device time for the initial GST experiment, and less than $0.2$ seconds per shot for the subsequent tune-ups. As such, this procedure could be incorporated into the daily calibration routines of these devices.}
The hybrid technique brings the gate as close as possible to its target, up to the hardware limit, but the actual effectiveness depends on the level of control afforded. \add{Basic calibration routines tend to be amplitude optimizations in Rabi experiments -- however the calibration routine in this particular instance is opaque to us. In expanding the calibration model to a full process matrix, it appears as though a logical tune-up to bring a quantum gate as close as possible to its target can be an effective and efficient calibration technique. Further, this procedure can detect and correct for unitary errors which do not add coherently, which can be a limitation of some detection schemes \cite{unitary-RB}.}
Although GST has previously been proposed as part of a quantum error mitigation protocol in \cite{GST-improvement-china} and \cite{Endo2018}, we emphasize the need to avoid repeated application of GST in any gate improvement techniques, owing to its extremely high experimental and computational overheads.\par 
We performed our experiments on an IBM cloud-based quantum computer with only logical level control. As a consequence, corrections to the CNOT could only be made through single qubit gate corrections, which themselves were erroneous. With control of the CNOT pulse scheme, \add{we anticipate that} the corrections \add{are likely to} be more effective.\par
In addition to the testing of the POST gate improvement scheme on a specific two-qubit case, we also conducted two-qubit GST experiments on six separate CNOT gates as an investigation into the performance and types of noise to occur on real superconducting devices. Understanding the real noise that occurs on devices is important for several reasons: It can help inform future characterization, which results in these procedures being less computationally expensive; it can help understand noise channels, which is important for quantum error correction \cite{OBrien2017}; and it can help identify hardware issues up on a real machine for better future implementation \cite{sandia-GST}. We present these results, as well as a theoretical evaluation of the effectiveness of our technique on the additional qubit pairs. 
\section{An overview of gate set tomography}
Gate set tomography is a hardware-agnostic method of characterizing quantum operations. This section provides a brief overview of its methodology -- for a comprehensive guide to the techniques involved, see \cite{intro-GST,sandia-GST}. In this work, we operate in the Pauli transfer matrix (PTM) representation of quantum channels. The matrices in this representation are mappings of the Stokes vector of a given density matrix. For some \add{channel} $\Lambda$, whose action on a density matrix $\rho$ has a Kraus decomposition $\Lambda\left(\rho\right) = \sum_i K_i\rho K_i^\dagger$, \add{its} matrix representation \add{$R_\Lambda$} is given by:
\begin{equation}
    (R_\Lambda)_{ij} = \text{Tr}\left(P_i\Lambda\left(P_j\right)\right),
\end{equation}
where the $P_{i,j}$ refer to the normalized, ordered set of Pauli matrices spanning the $d^2-$dimensional space of bounded linear operators on some Hilbert space $\mathscr{H}_d$. That is, the $n-$qubit basis is the set 
\begin{equation}
   \mathcal{P}_n = \{I/\sqrt{2}, X/\sqrt{2}, Y/\sqrt{2}, Z/\sqrt{2}\}^{\otimes n}.
\end{equation}
\add{The PTM representation is therefore a mapping of eigenvectors of the Pauli matrices.}
Applying the superoperator formalism makes this picture more convenient. Here, density matrices are represented as $d^2$-dimensional vectors $|\rho\rangle\rangle$ on a Hilbert-Schmidt space with inner product $\langle\langle\rho_1|\rho_2\rangle\rangle = \text{Tr}(\rho_1^\dagger\rho_2)$. This allows the action of quantum channels to be given by ordinary matrix multiplication. The $k$th component of these vectors is equal to $\text{Tr}(P_k\rho).$ The operational action is given by $|\Lambda(\rho)\rangle\rangle = R_\Lambda |\rho\rangle\rangle$ and map composition by $R_{\Lambda_2 \circ \Lambda_1} = R_{\Lambda_2}\cdot R_{\Lambda_1}$.\par
The only experimentally accessible quantities in a laboratory are \add{probabilities of measurement outcomes}. For example, after a sequence of quantum operations $G$, many measurements are taken in order to form an estimate of 
\begin{equation}
    \langle\langle E|G|\rho\rangle\rangle
\end{equation}

\begin{figure*}
    \centering
    \includegraphics[width=0.95\linewidth]{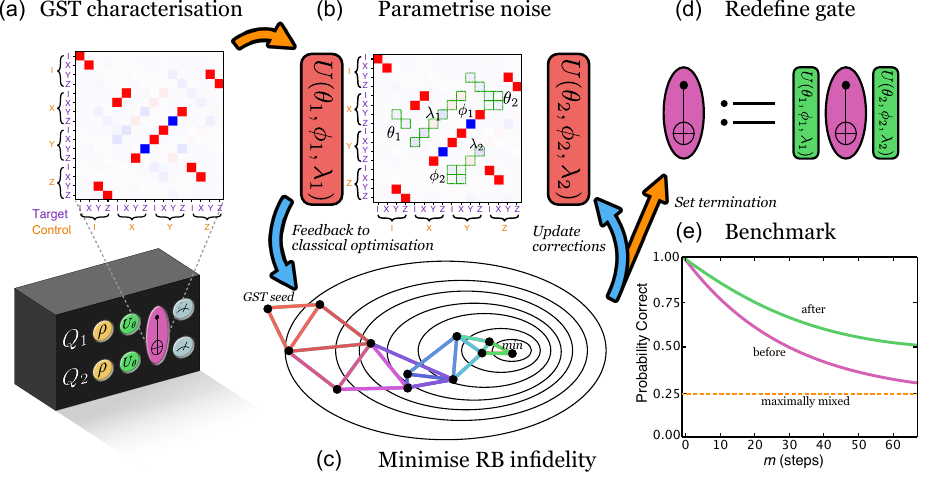}
    \caption{Overview of the POST approach for CNOT gate fidelity improvement. (a) Initial characterization of the CNOT Pauli transfer matrix (PTM) using gate set tomography. (b) Gate noise is parametrized in the action of arbitrary single-qubit unitaries acting before and after the CNOT gate. The parameter \add{values} that minimize the Frobenius distance between the noisy and ideal CNOT gates are then found. (c) These \add{values} then seed a Nelder-Mead optimization, where the objective function is the experimental infidelity of a chosen length randomized benchmarking experiment. With each iteration of POST, a new simplex is chosen and the randomized benchmarking experiment is performed for each vertex with the single qubit unitaries taking on the parameters at that point. (d) When the minimum infidelity is found, the CNOT gate is redefined to include the corrective unitaries. (e) The new gate is then fully benchmarked against the native gate, showing a revived fidelity for a much larger number of applications. }
    \label{flow-chart2}
\end{figure*}
for some preparation Hilbert-Schmidt vector $|\rho\rangle\rangle$ and some measurement effect $\langle\langle E|$. Quantum process tomography is a technique which provides an estimate for $G$ by acting the operator on a complete set of preparations, followed by a complete set of basis measurements. This style of characterization assumes \add{known} state preparation and measurement (SPAM). When SPAM errors are not negligible, however, QPT \add{can} produce gate estimates considerably far away from the true maps \cite{intro-GST}. This is particularly an issue since the primary source of error in current quantum computers are SPAM errors. RB curves \add{provide} a metric for the quality of a gate operation, but \add{this number alone does not elucidate} how that gate might be improved. Furthermore, since it is insensitive to SPAM errors, there is little or no information produced about the character of SPAM on a device. The ideal characterization should produce an accurate picture of all quantum channels, including projection operations. \par
GST aims to fully characterize a complete set of gates. The self-consistency in this method is achieved by including the preparation and measurement operations within the gateset $\{|\rho\rangle\rangle,\langle\langle E|,G_0,G_1,\cdots\}$. A set of gates is chosen firstly as the object of characterization. The only requirement is that these operations (or their compositions) generate an informationally complete set of preparations and measurements. That is, they form a complete basis of the Hilbert-Schmidt space. These SPAM operations are known as \emph{fiducials}, and are denoted by $\mathcal{F} = \{F_0,F_1,...,F_n\}$. They are optimally selected to form the most mutually distinguishable informationally complete set. Further, a set of gate compositions $\mathcal{G} = \{g_0,g_1,...,g_n\}$ is generated. The elements of this set are termed \emph{germs}, and each comprises a sequence of operations from the gateset. Gates contain a large number of free-parameters, and the emergence of errors in these depend on the input state, sequence of operations, and basis in which a measurement is made. Germs are chosen from an extensive search such that a possible error in each gate parameter may be amplified and made detectable by the repetition of at least one germ. In order to reduce the statistical error in detecting noise, each germ is repeated $L$ times for many different values of $L$. When every possible noisy parameter is made detectable, the germs set is termed \emph{amplificationally complete} \cite{sandia-GST}. For all values of $i,j,k,$ and $L$, the experimental data are then collected in the object 
\begin{equation}
    \label{GST_object}
    p_{ijk}^L = \langle\langle E| F^{(\text{prep})}_i g_j^L F^{(\text{meas})}_k|\rho\rangle\rangle.
\end{equation}

Linear inversion can provide an estimate of each gate at this point, but there is no natural way to include physicality constraints or to select the correct gauge. A maximum likelihood estimate (MLE) is therefore performed to provide the best estimate for the experimental gateset, consistent with (\ref{GST_object}), which obeys the conditions of complete positivity and trace preservation. \add{Following this is a final gauge optimization, selecting the gauge which most closely resembles the target gateset}.\par
GST provides a device-independent estimate of the quantum channel. Like all characterization processes, it assumes a model of the physical process taking place. The model assumption is purely Markovian, including zero leakage and weak environmental couplings -- that maps are composable and without context dependence. When violations of these model assumptions show up in the data, the non-Markovianity is then flagged. It is for this reason that the GST estimates must be treated with care: in physical terms, a quantum operation cannot necessarily be treated in isolation and a circuit decomposed into its constituent maps.\par

\section{Improvement of gate set parameters}
The CPTP map of quantum gates given by GST highlight all Markovian errors in the operation. Examples of errors of this nature include control errors, such as axis tilt or errors in pulse shaping, or erroneous coherent rotation of the qubits due to external couplings. \add{Our general goal in this work is to examine how best to feed this characterization back into a quantum device with less noise. We consider only the estimate determined for the CNOT gate}. The best method in which to address these noisy parameters depends on the level of control afforded in the device. Given the recent advent of cloud-based NISQ computers, where users only have restricted control of the device, the POST protocol introduced here makes use only of additional logical operations on the qubits (although we note in the conclusion that the extension to high-level pulse control is possible). At the time of the experiments, this was the only control available to the authors, and so the only scheme examined. 
\add{The natural CNOT gate on these devices is a cross-resonance $ZX$ interaction bookended by local rotations.}
With pulse level control, provided by IBM through OpenPulse \add{\cite{open-pulse}}, \add{we speculate that} the proposed blueprint could be modified to absorb the corrections into the \add{local pulses} -- rather than applying logical corrections around it. We summarize the overall POST procedure in Figure \ref{flow-chart2}, and describe it in further detail here.\par 
GST provides a means by which errors can be identified, but it is not necessarily straightforward to then mitigate their effects. Errors occurring on two-qubit gates such as a CNOT tend to be an order of magnitude greater than those of single qubits. From the perspective of logical corrections, it is therefore optimal to address two-qubit noise with the application of single-qubit gates. \par
Consider a quantum device with an informationally complete set of two-qubit controls. This can be used to conduct a GST experiment in the standard way on a qubit pair. The GST analysis of a CNOT produces an estimate for the CPTP map of the \add{gate}, designated by $\bar{G}_{CX}$. This can be decomposed into a the product of an ideal CNOT, $G_{CX}$, and some residual noise channel $G_\Lambda$ -- the inverse of which is generally unphysical \cite{inverseCP}. Previous approaches have typically treated this noise with quasiprobability decompositions \cite{Temme2017}. 
In the case of solely logical control, the nearest physical corrective $G_\Lambda^{-1}$ map may not be within the user's control-set. Importantly, given gate calibration and general hardware drift, a GST estimate's accuracy quickly expires over time. Solely utilising GST will require a significant overhead every time the corrections are implemented. From this, it is clear that GST on its own faces limitations as a practical method of improving gates.\par
\begin{figure*}
\centering
    \includegraphics[width=0.87\linewidth]{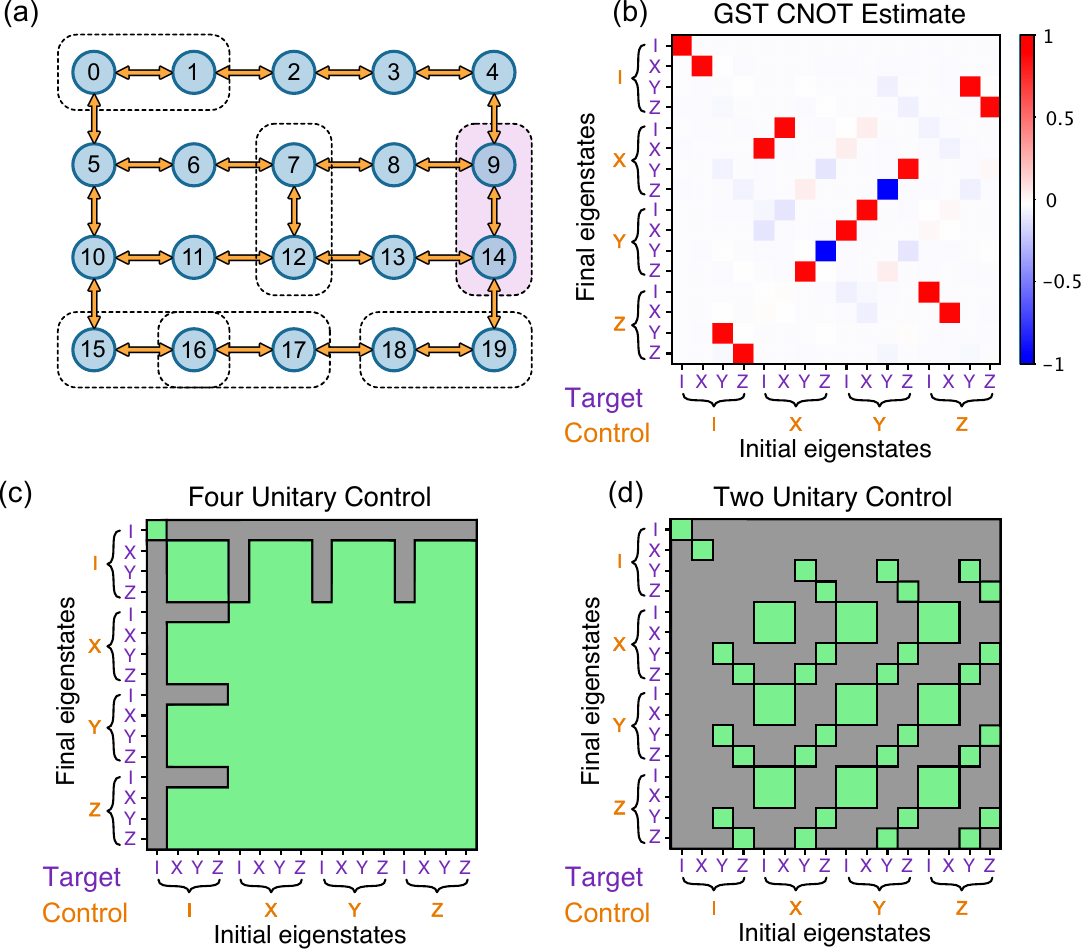}
    \caption{(a) The \emph{ibmq\_poughkeepsie} device layout, showing both the geometry and connectivity. Dotted lines indicate the six pairs of qubits that were characterized under GST. Qubits 9 and 14 (highlighted in pink) were chosen as the subject for testing POST. (b) The GST reconstructed estimate of the PTM of the 14-9 CNOT gate under investigation. (c) The green parts of this matrix indicate the controllable parts of the overall map when combining single-qubit operations on the control and target qubits both before and after the native CNOT. Because not all elements of the PTM can be \emph{independently} controlled, this does not imply that all error falling in the green can be corrected. Under this regime, almost all coherent noise is addressable in principle. (d) The green parts of this matrix indicate the controllable parts of the map when restricted to single-qubit operations on the control qubit before and after the native CNOT. This regime is far more restrictive than the previous for the trade-off of introducing approximately half of the amount of single-qubit error. These may also be viewed as indicators of the propagation of unitary error. }
    \label{noise-fig}
\end{figure*}

 
Without direct control of the hardware, correcting all \add{coherent} two-qubit errors will not be possible, since these corrections will contain associated errors equal or greater in magnitude than the existing ones. \add{Further, cross-resonance errors are not controllable at this level.} In this control regime we propose placing single qubit corrective gates before and after the native CNOT in order to correct as much of the local noise as possible, and then optimising over their parameters. Using a unitary parametrization for the four correction gates $U_i$ ($i\in\{1,2,3,4\}$)
\begin{equation}
\label{unitary-param}
    U_i(\theta_i,\phi_i,\lambda_i) = \begin{pmatrix}
\cos(\theta_i/2) & -e^{i\lambda_i}\sin(\theta_i/2) \\
e^{i\phi_i}\sin(\theta_i/2) & e^{i\lambda_i+i\phi_i}\cos(\theta_i/2) 
\end{pmatrix},
\end{equation}
We propose a super-logical CNOT gate structured as 
\begin{equation}
\label{twelve_param}
    (U_1\otimes U_2)\cdot \bar{G}_{CX} \cdot (U_3\otimes U_4),
\end{equation}
provided that the cumulative error of four single qubit gates is not greater than one two qubit gates. In the case of high single qubit error rates (or cross-talk between simultaneous gates), a similar approach can be made by applying local corrections exclusively on the control qubit \add{(this allows for broader manipulation than corrections solely on the target qubit)}, at the expense of more limited noise-targeting. That is,
\begin{equation}
\label{six_param}
    (U_1\otimes \mathbb{I})\cdot \bar{G}_{CX} \cdot (U_2\otimes \mathbb{I}).
\end{equation}
\add{The distinction between these two levels of correction is not immediately obvious.} In order to clearly see the difference between corrections on both qubits versus corrections acting solely on the control qubit, we illustrate the addressable parts of the CNOT matrix in green in Figures \ref{noise-fig}c and \ref{noise-fig}d. \add{That is, this indicates the extra accessible dimensions of control offered in either case. Whether that control is necessary in practice will depend on where the noise manifests and the trade-off in introducing further single-qubit error.} Note that that not all of elements in the green region can be necessarily independently configured. We call a matrix element `controllable' under a particular set of control operations if it is possible to change that matrix element by varying the parameters of the control. Matrix elements that are uncontrollable under a set of operations cannot be changed at all, and are marked in grey. Not all errors that fall in the green will necessarily be completely correctable. \par
The $\phi_i,\:\theta_i,\:\lambda_i$ are first selected with a simple optimization to minimize the distance between the corrected gate and the ideal map\add{. At this stage, any norm could be selected. For computational convenience, we chose the Frobenius norm, which also minimized the average gate infidelity,}
\begin{equation}
    \frob{U_1\otimes U_2 \cdot \bar{G}_{CX} \cdot U_3\otimes U_4 - G_{CX}}.
    \label{frob-dist}
\end{equation}
If the GST estimates of a quantum process were perfect and static with time, then this would be sufficient to have an improved CNOT gate. However, because GST is only an estimate of a Markovian map within a (generally) non-Markovian system, a simple mathematical minimization will not necessarily result in a physical optimization. \add{This could be either if the noisy parameter values were to change over time, or if the GST estimate were slightly incorrect.}
Instead, what we propose is a tune-up procedure which optimizes the \emph{performance} of the gate by using GST to identify \add{the neighbourhood in which certain parameter values are able to be modified to improve the accuracy of the gate.} As such, it is robust to the drift of different noisy parameters and does not rely on the absolute accuracy of the GST estimate. The only assumption is that the noisy \add{channel} will remain structurally similar enough to those of the GST estimate, that an optimization seeded by GST will bring us back to a better gate than the native operation in few iterations. We define our objective function as the RB infidelity of the gate, in order to make use of the most general metric of performance. \par 

The algorithm to implement the POST procedure is as follows:
\begin{enumerate}
    \item Conduct a series of experiments given by the requirements of GST. Use these to produce an output estimate of the gate's PTM. 
    \item Using a classical minimization technique, find the six or twelve parameter \add{values} which numerically minimize the Frobenius distance between the super-logical and the ideal CNOT gates, given in Expression (\ref{frob-dist}). These will be the seed parameter \add{values}. The choice of the Frobenius distance is not necessarily special, but we elected to use it to make the resulting matrices as similar as possible. 
    \item Define the objective function to be \add{the average survival probability over an ensemble of circuits for} some length $m$ RB experiment. Taking Step 2. as a newly defined CNOT gate, compute the objective function for both the native and new CNOT gates as a point of comparison. Using the parameters obtained from GST as a seed, perform an optimization of the CNOT gate by feeding the parameter \add{values} into the Nelder-Mead algorithm, where for each vertex of the simplex, the $m-$length RB infidelity is computed as the objective function.
    \item Converge at some pre-defined level of change. The newly improved CNOT gate is then defined by the composition of the final single qubit unitary gates on either side of the native CNOT gate.
    \item Conduct a full RB experiment to compare the new gate fidelity to the original. 
\end{enumerate}
The overview of the procedure is to use the GST estimate as a seed for the Nelder-Mead optimization algorithm, which then works to minimize the infidelity of the CNOT gate by varying the parameters given in (\ref{twelve_param}) or (\ref{six_param}). We take the infidelity measure to be a set of fixed $m$-length randomized benchmarking experiments. In a short number of iterations, this locates the optimal corrective rotations to make for a given day. This process is summarized in Figure \ref{flow-chart2}. The flexibility of the procedure is not only its robustness to drift, but generic steps (optimization procedure, noise parametrization, objective function) can all be chosen at the user's discretion. 
\add{Part of the rationale in performing GST is the ability to specifically determine the level of coherent noise in the gate from the outset, motivating whether it is possible to improve the gate at all.
Further, it provides useful information through the determination of the diamond distance between the noisy gate and the ideal.
The diamond distance is a means of assessing the distinguishability of two quantum channels. It is a worst-case error rate, taking the largest output trace distance over all possible input matrices. That is, for two quantum channels $\Phi_1$ and $\Phi_2$:
\begin{equation}
    ||\Phi_1-\Phi_2||_\Diamond := \sup_{\rho} \frac{1}{2}||\Phi_1\otimes \mathbb{I}_{d^2}(\rho) - \Phi_2\otimes \mathbb{I}_{d^2}(\rho)||_1,
\end{equation}
where $||\cdot||_1$ is the trace distance, a common measure of distinguishability between two density matrices. This metric between channels is commonly used in fault tolerance calculations for quantum error-correcting codes. It is also much more sensitive to coherent error in the gate than the average gate infidelity, which is typically the figure provided from RB curves. By performing GST, we are able to find the global minimum for the gate infidelity, which also globally minimizes the diamond distance. We discuss using our data in the next section how small differences in local minima of the infidelity can correspond to large changes in the diamond norm.}

\section{Experimental Implementation}
\begin{figure*}
\centering
    \includegraphics[width=0.73\linewidth]{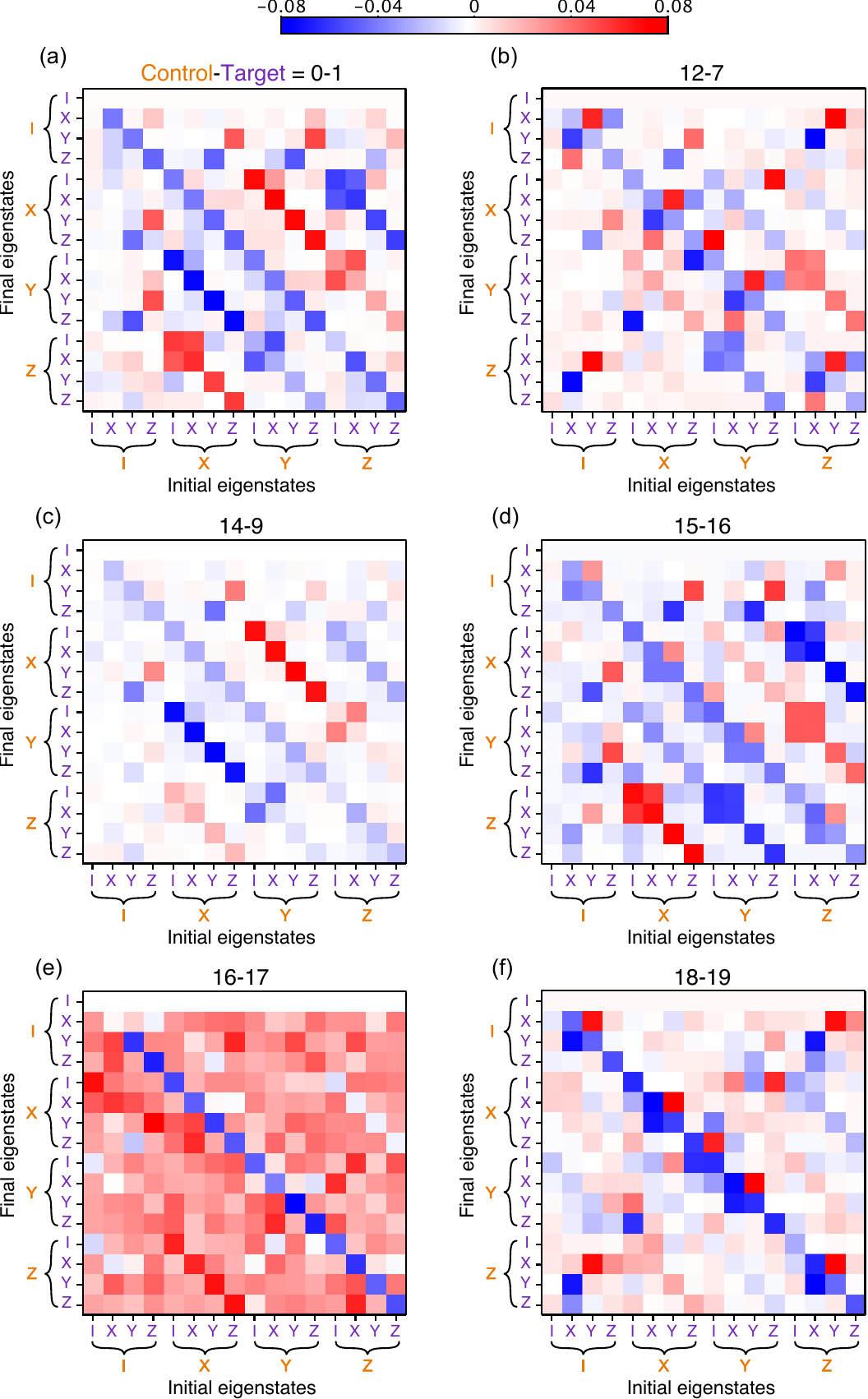}
    \caption{The matrices here show the GST-estimated PTM representations of the CNOT gate \add{error generator} between 6 pairs of different qubits. The control and target designations correspond to the device layout given in Figure \ref{noise-fig}. }
    \label{noise-ptms}
\end{figure*}

We tested the POST framework for CNOT characterization and improvement on the 20 qubit \emph{ibmq\_poughkeepsie} superconducting quantum device. Two qubit GST was performed on six pairs of qubits with the gateset $\{\mathbb{I},G_{XI}\left(\frac{\pi}{2}\right), G_{IX}\left(\frac{\pi}{2}\right),G_{YI}\left(\frac{\pi}{2}\right),G_{IY}\left(\frac{\pi}{2}\right),G_{CX}\}$ up to a germ repetition of $L=4$ for a total of $10\:500$ circuits at 8190 shots each. The layout and connectivity of this device is shown in Figure \ref{noise-fig}a. We also indicate the qubits on which experiments were performed. Using the notation `control-target' to indicate the physical qubit pair used respectively as the control and target of a CNOT gate, we characterized the gates of qubits 0-1, 12-7, 14-9, 15-16, 16-17, and 18-19. \add{In the estimation procedure, each gate was constrained to be CPTP.} We then elected to test the POST procedure on the gate which had most recently been characterized, for which the control was qubit $\#14$ and target qubit $\#9$. For the germ generation and MLE steps, we used the comprehensive open source Python package \emph{pyGSTi}, introduced in \cite{pygsti}. With the tools available, we generated the required germs from our target gatesets, and conducted the analysis of our experimental data. The GST estimate from the 14-9 qubits, used hereon in the POST tests, is shown in Figure \ref{noise-fig}b. The resulting noise maps for each additional CNOT gate can be found later in this manuscript, in Figure \ref{noise-ptms}. \add{In the gateset estimation procedure, \emph{pyGSTi} flags violation of the Markovian process matrix model, typically indicating some form of non-Markovianity in the device. Model violation is usually summarized in the form of difference between the observed and expected maximum log-likelihood $\log(\mathcal{L})$ for a given $k$ degrees of freedom. A model-abiding $2\Delta\log(\mathcal{L})$ is expected to be distributed with mean $k$ and standard deviation $\sqrt{2k}$. For length-1 sequences where $k=737$, our data showed $2\Delta\log(\mathcal{L}) = 2861$, for length-2 where $k=6006$, $2\Delta\log(\mathcal{L}) = 15\:938$, and for length-4 where $k=26\:901$, $2\Delta\log(\mathcal{L}) = 107\:889$. This suggests that the device was not totally Markovian, and the estimation not totally perfect.}

\begin{figure*}
\centering
    \includegraphics[width=0.89\linewidth]{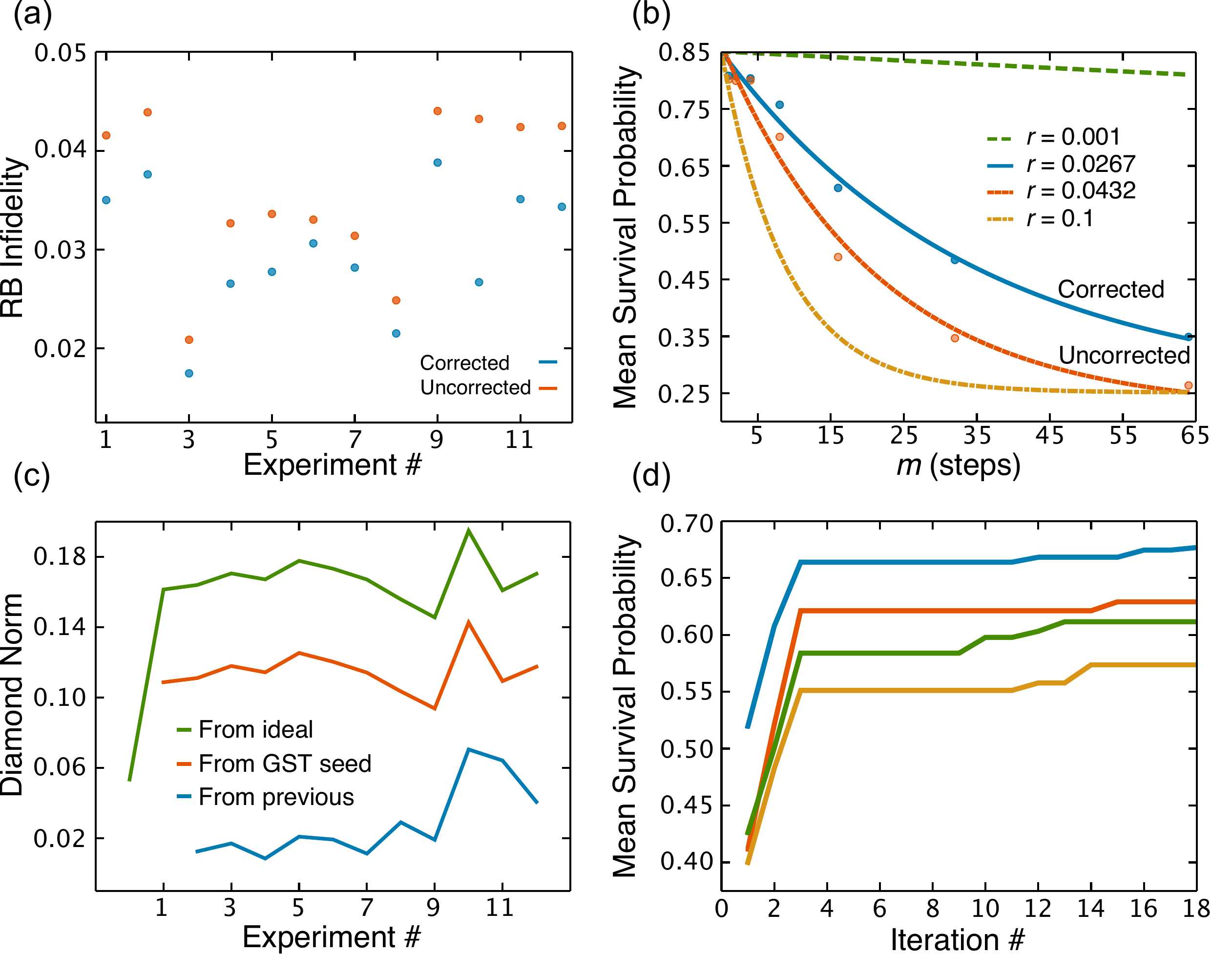}
    \caption{Efficacy and robustness of the POST protocol -- experimental test on the 14-9 CNOT gate. (a) \add{RB values for each experiment taken over the six week period, both for the corrected and the uncorrected gates.} (b) Example RB curve showing our best improved CNOT gate decay vs. the native gate RB experiment for experiment 10 (our most emphatic example of gate improvement). Also shown for comparison are theoretical curves corresponding to 10\% and 0.1\% error rates. (c) A comparison of the diamond norm of the corrective channel over time: firstly with respect to the ideal CNOT, secondly with respect to the initial GST corrections, and finally with respect to each previous experiment. (d) An indication of the convergence speed of the Nelder-Mead gate optimizer for four example runs. This figure shows how the  \add{survival probability} of the RB experiment increases with each iteration. We observe an initially large increase in the fidelity of the experiment, followed by only small changes thereafter, and zero changes after 18 iterations. The sequence very quickly finds its best gate in a small number of steps.}
    \label{rb_curve}
\end{figure*}

\begin{table*}[]
    \centering
\begin{tabular}{@{}lp{1.1cm}p{1.1cm}p{1.1cm}p{1.1cm}p{1.1cm}p{1.1cm}@{}}
\toprule
Date & $\theta_1$ & $\phi_1$ & $\lambda_1$ & $\theta_2$ & $\phi_2$ & $\lambda_2$ \\ \midrule

31/03/19 (initial GST) & 0.046 & -1.271 & 0.480 & 0.029 & 0.480 & 0.393 \\
02/04/19 & 0.116 & -1.234 & 0.536 & 0.116 & 0.545 & 0.403 \\
09/04/19 & 0.130 & -1.218 & 0.461 & 0.084 & 0.565 & 0.475 \\
24/04/19 & 0.116 & -1.234 & 0.552 & 0.119 & 0.558 & 0.415 \\
27/04/19 & 0.148 & -1.208 & 0.502 & 0.010 & 0.530 & 0.427 \\
29/04/19 & 0.089 & -1.228 & 0.523 & 0.071 & 0.523 & 0.436 \\
01/05/19 & 0.218 & -1.199 & 0.552 & 0.010 & 0.552 & 0.365 \\
03/05/19 & 0.089 & -1.228 & 0.523 & 0.071 & 0.523 & 0.436 \\
10/05/19 & 0.096 & -1.221 & 0.530 & 0.079 & 0.530 & 0.443 \\
13/05/19 & 0.141 & -1.310 & 0.497 & 0.123 & 0.575 & 0.488 \\* \bottomrule

\end{tabular}
\caption{Values of the corrective parameters obtained after a full improvement cycle for each day of experiments, given to the third decimal place. These correspond to the unitary parametrization given in Equation (\ref{unitary-param}).}
\label{param-table}
\end{table*}
\emph{Results Summary} -- The initial GST analysis of the 14-9 CNOT gate took place on the 31st of March, 2019 and its corrective parameters used as the base vertex for the Nelder-Mead simplex method. The procedure was implemented a total of 12 times over a period of approximately six weeks, corresponding to overlap with approximately 40 different calibration cycles. Figure \ref{rb_curve}a displays a summary of the improvement shown over the native gate with each experiment run for bare RB infidelity $r_{\rm u}$ and corrected RB infidelity $r_{\rm c}$. In each case, both the corrected and uncorrected benchmarking experiments were conducted in the same job submission to avoid any bias in gate drift throughout the day\add{, and the tune-up was often conducted shortly after the calibration of the machines}. The total average improvement\add{, which we define as $r_{\rm u}/r_{\rm c} - 1$} was $21.1\%$, with a notable outlier of $61.8\%$ in experiment 10. The median observed improvement was $19.1\%$. In the next section we discuss how this compares to theoretical figures based on the GST estimates. Figure \ref{rb_curve}b is a comparison RB curve showing the decay of an example improved gate over the native fidelity. For clarity and comparison, we also plot example curves with $10\%$ and $0.1\%$ error rates. Note that this RB number is from the overall curve, which is composed of single and two-qubit gates. For these experiments, this partitions into $r = 3/4\cdot r_{\rm CNOT} + 1/4\cdot r_{\rm single}$. To reduce the total number of experiments per day, we did not compute multiple curves with different fractions of CNOT and single qubit gates. Consequently, $r_{\rm u} /r_{\rm c} - 1$ is really $(3/4\cdot r_{\rm u,CNOT}+1/4\cdot r_{\rm single})/(3/4 \cdot r_{\rm c,CNOT}+1/4\cdot r_{\rm single}) - 1$, which is a lower bound for the improvement of the CNOT gate. Given that $r_{\rm single}< r_{\rm CNOT}$ by about an order of magnitude, we do not expect that the figure differs substantially.\par 
The minimized objective function was the average infidelity of 20 randomly sampled RB circuits, consisting of 16 circuit layers in addition to the preparation and measurement layers. Each circuit was run at 8190 shots to minimize statistical error in the optimization. The use of an RB experiment as the objective function is a flexible metric and can be chosen as the user desires. In principle, context-dependence of a gate may affect the versatility of the improved gate \add{in the sense that gates may affect their system differently depending on the circuit}, however at this stage RB curves are the most robust assessment of a gate's performance and require the fewest assumptions. \par 
\add{Although inserting four local unitaries provides more coherent control, we found empirically on this device that this introduced more error than it eliminated. Consequently, we chose to apply corrections only on the control qubit, before and after the CNOT gate. This reduces the optimization to 6 parameters.}
In the Nelder-Mead method, a dimension 6 simplex with 7 vertices is constructed, with the base vertex given by the GST parameters. The objective function is then evaluated for each point. In order to save on computation, we elected to omit the shrink step. After five iterations of no further improvement we would then terminate the algorithm and redefine our gate with the best point. \par
We used a relatively new form of RB known as \emph{direct randomized benchmarking} (DRB) \cite{DRB}. In a single circuit, DRB prepares stabilizer states, followed by $m$ randomly selected layers of gates native to that stabilizer, before finally performing a stabilizer measurement to give the success probability. A number of randomly generated circuits can then be used to provide an overall average. The utility of this over Clifford RB is the ability to specify the occurrence of given gates. \add{It also accommodates for future instances of the protocol where the tune-up might be efficiently computed over a larger number of qubits.} Here, we randomly generated 20 RB circuits, with CNOT gates composing on average $3/4$ of the total circuit. The average probability of success at length $m$, $P_m$
is then plotted over a series of values for $m$. These points are then fit to $P_m = A + Bp^m$ for fit parameters $A,B$, and $p$. The RB number $r = (15/16)(1-p)$ is then the probability of an error occurring under a stochastic model. \add{We remark here that the RB number is not necessarily a true estimate of the average gate infidelity and can differ under, for example, gate-dependent noise. Here, our characterization produced an initial infidelity estimate of $1.191\times10^{-2}$ for the CNOT gate, whereas the estimated RB number was $3.369\times10^{-2}$. RB is currently the most efficient method for estimating the quality of a gate, and the number most often cited in reporting the performance of hardware. For this reason, we use this as indicative of the gate infidelity, even though the true figures may differ. Indeed, this suggests that IBM may have better gates than their RB curves suggest. For further details, see } \cite{PhysRevLett.119.130502,noise-coherence-2015}. \par
\add{As previously stated, the utility of the GST experiments is both in determining the coherence of the gate noise and in effecting a search which is placed near the global minimum of both gate infidelity and the diamond norm.
For example, using the GST estimate we simulated the randomized benchmarking tune-up when seeded from the zero vector (that is, the case in which we optimize the RB number without the initial GST seed). In this case, the Nelder-Mead search terminates at a point where $r_{\text{CNOT}} = 0.0254$ and $\mathcal{E}_\Diamond=0.0426$ -- whereas the numerical search for the global minimum yields $r_{\text{CNOT}} = 0.0210$ and $\mathcal{E}_\Diamond = 0.0286$. 
Randomly seeded starts were able to find the global minimum, but only with the equivalent of between $10\:000$ and $20\:000$ RB experiments, which is comparable to the initial number of circuits required for GST. Crucially, this approach comes with no evidence of having reached the global minimum, which has large implications for $\mathcal{E}_\Diamond$.}
%\com{Follow up or add in a paragraph about the utility of the GST seed in terms of simulation with the figures that I gave Lloyd and Charles.}
%\com{Whilst we can’t run the experiments for RB optimization with the zero seed, I can at the very least simulate it out with the GST estimate. In this case, the Nelder-Mead gets stuck in a local minimum where the RB number is 0.0254 (vs. 0.0210) and the diamond norm is 0.0426 (vs. 2.86). If I run with random starting points, then this can reach the global minimum in about 1000 function evaluations (~10000-20000 RB circuits) which is comparable to the initial GST anyway. Crucially, you can’t know if you’re in a local minimum or not with this method, and as you can see, this also plays a big role on the diamond norm.}
\par
\add{\emph{Tracking the tune-up}} --
\add{It is prudent to ask how much the corrective values changed over time. We stop short of claiming that these values directly quantify drift, since a flat objective function could plausibly entail corrections whose value changed without much impact on the RB number. However, empirically we found that the previous day's optimal corrections would typically reduce the performance of the gate, rather than improving it. It was necessary therefore to perform the optimization, rather than simply applying the results of previous days.}
Operating on the assumption that this method finds the most appropriate \add{values for the} corrective parameters on a particular day, we can use this data to loosely quantify the change in the coherent noise on a real quantum information processor (QIP). \add{The values of the corrective parameters for each experiment are provided in Table \ref{param-table}}. Moreover, the variables do not all independently affect the final map, meaning that change in the parameter \add{values} themselves might obfuscate the fact that the channel overall has not varied much. \add{In this table, the $\theta$ values appear to change the most. These values are equivalent to $Y-$rotations, which change the populations of the control qubit. The reason for this is not entirely clear, however, inspection of the error channels in Figure \ref{noise-ptms}c shows that some of the most dominant noise occurs in the $X$ and $Z$ control qubit blocks. This could mean that they are more likely to change with time.} \par 
In order to paint a more concrete picture of the effects of the parameters, we study the case of the improved gate as though it comprised three perfect gates. We then compare the diamond distance of this channel from the ideal CNOT, from the initial GST seed, and from the previous experiment. Our results are summarized in Figure \ref{rb_curve}c. One of the key assumptions in this method was that the system and its gate noise never changed too much from the initial GST seed that the optimizer could not easily find the best gate for the day. The distance of the corrective channel on a given day from the GST seed supports this stance: over a six week period the mean diamond distance is 0.115, with a standard deviation of $1.2\times10^{-2}$.\par 
\emph{Convergence Speed} -- Any error mitigation protocol designed to address drift in a QIP will need to be regularly implemented as part of a tune-up procedure. It is therefore ideal that it require as few experiments as possible. In Figure \ref{rb_curve}d we present how each iteration of the Nelder-Mead optimization increased the \add{survival probability} of the RB experiment for four sample experiments. In each case, a substantial improvement was found in the first three iterations, beyond which we observed only small fluctuations, and zero change after 18 iterations. Each iteration requires 11 circuits to run, and so most of the improvement was found in 22 circuits, with the worst case requiring approximately 200 -- depending on when one accepts the algorithm to terminate. Assuming 0.5 milliseconds per shot, POST would take approximately nine minutes to converge in the worst case scenario. Depending on the cross-talk limitations of the device, it could then be run in parallel across all non-overlapping qubit pairs.

\begin{table*}[]
\centering
%\begin{tabular}{@{}cccccc@{}}
\begin{tabular}{@{}lllllll@{}}
\toprule 
%\begin{tabular}[c]{@{}l@{}}Qubit pair \\ (control-target)\end{tabular} & $r_{\rm CNOT}$ & $\mathcal{E}_{\Diamond \rm CNOT}$ & \begin{tabular}[c]{@{}l@{}}Six parameter \\ minimum $r_{\rm CNOT}$\end{tabular} & \begin{tabular}[c]{@{}l@{}}Six parameter minimum \\ $\mathcal{E}_{\Diamond \rm CNOT}$\end{tabular} & \begin{tabular}[c]{@{}l@{}}Coherent minimum \\ $r_{\rm CNOT}$\end{tabular} & \begin{tabular}[c]{@{}l@{}}Coherent minimum\\ $\mathcal{E}_{\Diamond \rm CNOT}$\end{tabular} \\ \midrule
%0-1 & $2.283\times 10^{-2}$ & $4.739\times 10^{-2}$ & $1.785\times 10^{-2}$ & $2.435\times 10^{-2}$ & $1.777\times 10^{-2}$ & $2.259\times 10^{-2}$ \\
%12-7 & $4.515\times 10^{-2}$ & $9.022\times 10^{-2}$ & $4.164\times 10^{-2}$ & $7.770\times 10^{-2}$ & $2.880\times 10^{-2}$ & $3.334\times 10^{-2}$ \\
%\rowcolor[HTML]{DAE8FC} 
%14-9 & $3.369\times 10^{-2}$ & $6.827\times 10^{-2}$ & $2.102\times 10^{-2}$ & $2.860\times 10^{-2}$ & $2.071\times 10^{-2}$ & $2.740\times 10^{-2}$ \\
%15-16 & $3.545\times 10^{-2}$ & $8.099\times 10^{-2}$ & $2.293\times 10^{-2}$ & $3.786\times 10^{-2}$ & $2.213\times 10^{-2}$ & $2.787\times 10^{-2}$ \\
%16-17 & $5.558\times 10^{-2}$ & $8.092\times 10^{-2}$ & $5.258\times 10^{-2}$ & $7.703\times 10^{-2}$ & $5.093\times 10^{-2}$ & $7.070\times 10^{-2}$ \\
%18-19 & $2.197\times 10^{-2}$ & $4.392\times 10^{-2}$ & $2.163\times 10^{-2}$ & $4.210\times 10^{-2}$ & $1.944\times 10^{-2}$ & $3.110\times 10^{-2}$ \\ \bottomrule
%\end{tabular}
\begin{tabular}[c]{@{}l@{}}Qubit pair \\ (control-target)\end{tabular} & $r_{\rm CNOT}$ & $\mathcal{E}_{\Diamond \rm CNOT}$ & \begin{tabular}[c]{@{}l@{}}POST \\ minimum $r_{\rm CNOT}$\end{tabular} & \begin{tabular}[c]{@{}l@{}}POST \\minimum $\mathcal{E}_{\Diamond \rm CNOT}$\end{tabular} & \begin{tabular}[c]{@{}l@{}} Stochastic $r_{\rm CNOT}$\end{tabular} & \\ \midrule
0-1 & $2.28\times 10^{-2}$ & $4.74\times 10^{-2}$ & $1.79\times 10^{-2}$ & $2.44\times 10^{-2}$ & $1.78\times 10^{-2}$ \\
12-7 & $4.52\times 10^{-2}$ & $9.02\times 10^{-2}$ & $4.16\times 10^{-2}$ & $7.77\times 10^{-2}$ & $2.88\times 10^{-2}$ \\
\rowcolor[HTML]{DAE8FC} 
14-9 & $3.37\times 10^{-2}$ & $6.83\times 10^{-2}$ & $2.10\times 10^{-2}$ & $2.86\times 10^{-2}$ & $2.07\times 10^{-2}$ \\
15-16 & $3.55\times 10^{-2}$ & $8.10\times 10^{-2}$ & $2.29\times 10^{-2}$ & $3.79\times 10^{-2}$ & $2.21\times 10^{-2}$  \\
16-17 & $5.56\times 10^{-2}$ & $8.09\times 10^{-2}$ & $5.26\times 10^{-2}$ & $7.70\times 10^{-2}$ & $5.09\times 10^{-2}$  \\
18-19 & $2.20\times 10^{-2}$ & $4.39\times 10^{-2}$ & $2.16\times 10^{-2}$ & $4.21\times 10^{-2}$ & $1.94\times 10^{-2}$ \\ \bottomrule
\end{tabular}
\caption{
\add{For each GST analysis of the qubit pairs, we present the RB number of the CNOT gate, as well as its diamond distance from the ideal CNOT. Further, we show the same figures in the idealized scenario where perfect single qubit corrective unitaries are employed on the control qubit to remove coherent error. In the final column, we have included the predicted RB number for each gate with the complete removal of coherent noise. The diamond norm of almost every entry stands to be significantly improved with the removal of coherent error.}}
\label{tab:theoretical-improvements}
\end{table*}

\subsection{Tomography on other qubit pairs}
In addition to gate improvement on qubits 14 and 9 on the \emph{ibmq\_poughkeepsie}, we also performed GST experiments on five other qubit CNOT pairs on the quantum device. This incorporates a further 5 experiments of between $9367$ and $20\:530$ circuits at 8190 shots each. We present this data as a case study for the noise that occurs in a real quantum device, and theoretically assess the effectiveness of the POST technique at mitigating the noise. Here, the fidelities of these gates and their structure indicate whether noise can be addressed by the single-qubit unitaries used. Figure \ref{noise-ptms} shows the noise PTMs for each of the six CNOT gates investigated. \add{Specifically, we present this in the form of error generators $\mathcal{L}$ such that $\bar{G}_{CX}$ = $\text{e}^{\mathcal{L}} G_{CX}$. This presents the noise as though it were to occur after the ideal gate which allows for a helpful -- though not necessarily physically accurate -- picture.} The control and target numbers given refer respectively to the qubits of Figure \ref{noise-fig}a acting as the control and target of the CNOT gate under characterization. We elected to map out these qubits in order to obtain a relatively uniform sample of the full device geometry. It is instructive to compare these matrix plots with the schematics given in Figures \ref{noise-fig}c and \ref{noise-fig}d, which respectively indicate local target/control, and sole control rotations. PTM noise whose locations are correspondingly indicated in green in the schematics can be explained as a local rotation occurring either before or after the CNOT. For example, the block-like features prominent in 0-1 and 15-16 make up the landscape of Figure \ref{noise-fig}d, suggesting a rotation of $Z-$eigenstates of the control qubit into $X-$ and $Y-$ eigenstates both before and after the CNOT. Any noise that falls outside the green of either schematic can be attributed either to decoherence or cross-resonance errors.\par
We also use this data to estimate the effectiveness of the POST procedure on the other qubit-pairs. \add{Using the GST estimates for each CNOT gate, we compute its RB number as well as its diamond distance to the ideal case. We then investigate how these numbers are ideally minimized both in the case with corrective unitaries on the target qubit, and where all coherent noise is removed. This provides an insight into the location of most of the noise in these gates, as well as a partition of the gate infidelity into coherent and stochastic error. Interestingly, in most cases (barring the 12-7 pair) there is little difference between the best ideal gate using two single-qubit corrective unitaries, and where a perfect $U(4)$ gate is used to remove all coherent noise.} A summary of the data is provided in Table \ref{tab:theoretical-improvements}. \par 
Pulse sequences for the implementation of a CNOT gate typically consist of a local pulse to each qubit, as well as an additional cross-resonance pulse coupling the two. We would expect implementing POST with absorbed corrective rotations into the native CNOT pulse sequence would see a much larger increase in fidelity.
\section{Discussion}
The transition from mathematical maps to physical operations is not always a seamless one. Besides errors in the GST characterization, absent a good method of characterizing non-Markovian behaviour, assumptions must be made of weak system-environment correlations, composability of operations, and minimal cross-talk between qubits. The emergence of unexpected behaviour from quantum systems means that in-principle operational improvements, such as the direct application of corrections from GST estimates, cannot always be relied upon. We have presented a general quantum-classical hybrid method which uses the real-life performance of the gate as the feed-forward for corrective updates. The success of the procedure is therefore self-fulfilling. \par 
randomized benchmarking is a robust method of measuring the Markovian fidelity of a given operation. However, in a system with environmental back-action or context dependent gates, it is not clear whether the situation will always be so simple as transplanting a redefined gate into a quantum circuit and seeing an increased fidelity in this new context. The markedly better performance of quantum algorithms consisting of these redefined gates remains to be demonstrated and will be the subject of future work. In particular, the POST algorithm would be easily adapted to any characterization technique more inclusive of non-Markovian behaviour. \par
Developing high-fidelity gate hardware is imperative for the field of quantum computing to achieve its ambitious aims. An underrated measure of device quality, however, is \emph{consistency} -- the ability to achieve reported minimal error rates again and again despite gradual changes in device parameters and system-environment correlations. In this work we presented a consistent method that combines an initial overhead of gate set tomography with a classical optimization algorithm that delivers an improved two-qubit gate in relatively few experiments. We emphasize that although POST was tested on an \emph{IBM Quantum} device, it is applicable to any hardware with logical-level control. Furthermore, the method is adaptable to any level of control. The key aspect is identifying noisy parameters from the GST estimate using the afforded set of device controls. In particular, we would expect to see significantly better results with pulse-level control wherein instead of separately implementing the corrective unitaries, they would be absorbed into modifying the CNOT pulse, there would be minimal additional gate errors introduced, or increase in depth. 
\section{Acknowledgments}
We are grateful to K. Modi and F. Pollock for valuable conversations, and to D. Broadway for figure advice. \add{This work was supported by the University of Melbourne through the establishment of an IBM Quantum Network Hub at the University. 
GALW is supported by an Australian Government Research Training Program Scholarship. 
CDH is supported through a Laby Foundation grant at The University of Melbourne. Computational resources are acknowledged from the National Computational Infrastructure (NCI) and Pawsey Supercomputer Center under the National Computational Merit based Allocation Scheme (NCMAS).}

\section{references}


%merlin.mbs apsrev4-1.bst 2010-07-25 4.21a (PWD, AO, DPC) hacked
%Control: key (0)
%Control: author (8) initials jnrlst
%Control: editor formatted (1) identically to author
%Control: production of article title (-1) disabled
%Control: page (0) single
%Control: year (1) truncated
%Control: production of eprint (0) enabled
\begin{thebibliography}{0}%
\makeatletter
\providecommand \@ifxundefined [1]{%
 \@ifx{#1\undefined}
}%
\providecommand \@ifnum [1]{%
 \ifnum #1\expandafter \@firstoftwo
 \else \expandafter \@secondoftwo
 \fi
}%
\providecommand \@ifx [1]{%
 \ifx #1\expandafter \@firstoftwo
 \else \expandafter \@secondoftwo
 \fi
}%
\providecommand \natexlab [1]{#1}%
\providecommand \enquote  [1]{``#1''}%
\providecommand \bibnamefont  [1]{#1}%
\providecommand \bibfnamefont [1]{#1}%
\providecommand \citenamefont [1]{#1}%
\providecommand \href@noop [0]{\@secondoftwo}%
\providecommand \href [0]{\begingroup \@sanitize@url \@href}%
\providecommand \@href[1]{\@@startlink{#1}\@@href}%
\providecommand \@@href[1]{\endgroup#1\@@endlink}%
\providecommand \@sanitize@url [0]{\catcode `\\12\catcode `\$12\catcode
  `\&12\catcode `\#12\catcode `\^12\catcode `\_12\catcode `\%12\relax}%
\providecommand \@@startlink[1]{}%
\providecommand \@@endlink[0]{}%
\providecommand \url  [0]{\begingroup\@sanitize@url \@url }%
\providecommand \@url [1]{\endgroup\@href {#1}{\urlprefix }}%
\providecommand \urlprefix  [0]{URL }%
\providecommand \Eprint [0]{\href }%
\providecommand \doibase [0]{http://dx.doi.org/}%
\providecommand \selectlanguage [0]{\@gobble}%
\providecommand \bibinfo  [0]{\@secondoftwo}%
\providecommand \bibfield  [0]{\@secondoftwo}%
\providecommand \translation [1]{[#1]}%
\providecommand \BibitemOpen [0]{}%
\providecommand \bibitemStop [0]{}%
\providecommand \bibitemNoStop [0]{.\EOS\space}%
\providecommand \EOS [0]{\spacefactor3000\relax}%
\providecommand \BibitemShut  [1]{\csname bibitem#1\endcsname}%
\let\auto@bib@innerbib\@empty
%</preamble>
\end{thebibliography}%


\begin{thebibliography}{10}

\bibitem{sandia-GST}
R.~Blume-Kohout, J.~K. Gamble, E.~Nielsen, K.~Rudinger, J.~Mizrahi, K.~Fortier,
  and P.~Maunz.
\newblock {Demonstration of qubit operations below a rigorous fault tolerance
  threshold with gate set tomography}.
\newblock {\em Nature Communications}, 8:1--13, 2017.

\bibitem{Arute2019}
F.~Arute et~al.
\newblock {Quantum supremacy using a programmable superconducting processor}.
\newblock {\em Nature}, 574(July), 2019.

\bibitem{IBM-53}
C.~Nay.
\newblock {IBM Opens Quantum Computation Center in New York; Brings World's
  Largest Fleet of Quantum Computing Systems Online, Unveils New 53-Qubit
  Quantum System for Broad Use}, 2019.

\bibitem{intel-49}
J.~Hsu.
\newblock {Intel's 49-Qubit Chip Shoots for Quantum Supremacy}, 2018.

\bibitem{rigetti-19}
W.~Zeng.
\newblock {Unsupervised Machine Learning on Rigetti 19Q with Forest 1.2}, 2017.

\bibitem{Barends2014}
R.~Barends, J.~Kelly, A.~Megrant, A.~Veitia, D.~Sank, E.~Jeffrey, T.~C. White,
  J.~Mutus, A.~G. Fowler, B.~Campbell, et~al.
\newblock {Superconducting quantum circuits at the surface code threshold for
  fault tolerance}.
\newblock {\em Nature}, 508:500, 2014.

\bibitem{Zeuner2018}
J.~Zeuner, A.~N. Sharma, M.~Tillmann, R.~Heilmann, M.~Gr{\"{a}}fe, A.~Moqanaki,
  A.~Szameit, and P.~Walther.
\newblock {Integrated-optics heralded controlled-NOT gate for
  polarization-encoded qubits}.
\newblock {\em npj Quantum Information}, 4(1), 2018.

\bibitem{PhysRevLett.117.140501}
T.~P. Harty, M.~A. Sepiol, D.~T.~C. Allcock, C.~J. Ballance, J.~E. Tarlton, and
  D.~M. Lucas.
\newblock {High-Fidelity Trapped-Ion Quantum Logic Using Near-Field
  Microwaves}.
\newblock {\em Phys. Rev. Lett.}, 117(14):140501, 9 2016.

\bibitem{He2019}
Y.~He, S.~K. Gorman, D.~Keith, L.~Kranz, J.~G. Keizer, and M.~Y. Simmons.
\newblock {A two-qubit gate between phosphorus donor electrons in silicon}.
\newblock {\em Nature}, 571(7765):371--375, 2019.

\bibitem{Hong2019}
S.~S. Hong, A.~T. Papageorge, P.~Sivarajah, G.~Crossman, N.~Didier, A.~M.
  Polloreno, E.~A. Sete, S.~W. Turkowski, M.~P. da~Silva, and B.~R. Johnson.
\newblock Demonstration of a parametrically activated entangling gate protected
  from flux noise.
\newblock {\em Phys. Rev. A}, 101:012302, 2020.

\bibitem{Kjaergaard2019}
M.~Kjaergaard, M.~E. Schwartz, J.~Braumüller, P.~Krantz, J.~I.-J. Wang,
  S.~Gustavsson, and W.~D. Oliver.
\newblock Superconducting qubits: Current state of play.
\newblock {\em Annual Review of Condensed Matter Physics}, 11(1):369--395,
  2020.

\bibitem{Bradley2019}
C.~E. Bradley, J.~Randall, M.~H. Abobeih, R.~C. Berrevoets, M.~J. Degen, M.~A.
  Bakker, M.~Markham, D.~J. Twitchen, and T.~H. Taminiau.
\newblock {A Ten-Qubit Solid-State Spin Register with Quantum Memory up to One
  Minute}.
\newblock {\em Physical Review X}, 9(3):31045, 2019.

\bibitem{Klimov2018}
P.~V. Klimov, J.~Kelly, Z.~Chen, M.~Neeley, A.~Megrant, B.~Burkett, R.~Barends,
  K.~Arya, B.~Chiaro, Yu~Chen, et~al.
\newblock Fluctuations of energy-relaxation times in superconducting qubits.
\newblock {\em Phys. Rev. Lett.}, 121:090502, 2018.

\bibitem{Fogarty2015}
M.~A. Fogarty, M.~Veldhorst, R.~Harper, C.~H. Yang, S.~D. Bartlett, S.~T.
  Flammia, and A.~S. Dzurak.
\newblock {Nonexponential fidelity decay in randomized benchmarking with
  low-frequency noise}.
\newblock {\em Physical Review A - Atomic, Molecular, and Optical Physics},
  92(2):1--7, 2015.

\bibitem{Chow2009}
J.~M. Chow, J.~M. Gambetta, L.~Tornberg, Jens Koch, Lev~S. Bishop, A.~A. Houck,
  B.~R. Johnson, L.~Frunzio, S.~M. Girvin, and R.~J. Schoelkopf.
\newblock {Randomized benchmarking and process tomography for gate errors in a
  solid-state qubit}.
\newblock {\em Physical Review Letters}, 102(9):1--4, 2009.

\bibitem{PhysRevLett.78.390}
J.~F. Poyatos, J.~I. Cirac, and P.~Zoller.
\newblock {Complete Characterization of a Quantum Process: The Two-Bit Quantum
  Gate}.
\newblock {\em Phys. Rev. Lett.}, 78(2):390--393, 1 1997.

\bibitem{PhysRevA.77.012307}
E.~Knill, D.~Leibfried, R.~Reichle, J.~Britton, R.~B. Blakestad, J.~D. Jost,
  C.~Langer, R.~Ozeri, S.~Seidelin, and D.~J. Wineland.
\newblock {Randomized benchmarking of quantum gates}.
\newblock {\em Physical Review A - Atomic, Molecular, and Optical Physics},
  77(1):12307, 1 2008.

\bibitem{gst-2013}
R.~Blume-Kohout, J.~Gamble, E.~Nielsen, J.~Mizrahi, J.~Sterk, and P.~Maunz.
\newblock {Robust, self-consistent, closed-form tomography of quantum logic
  gates on a trapped ion qubit}.
\newblock {\em arXiv:1310.4492}, 2013.

\bibitem{PhysRevA.87.062119}
S.~T. Merkel, J.~M. Gambetta, J.~A. Smolin, S.~Poletto, A.~D. C{\'{o}}rcoles,
  B.~R. Johnson, C.~A. Ryan, and M.~Steffen.
\newblock {Self-consistent quantum process tomography}.
\newblock {\em Phys. Rev. A}, 87(6):62119, 6 2013.

\bibitem{osti_1428158}
P.~L.~W. Maunz.
\newblock {Characterization of Two-Qubit Quantum Gates in Sandia's High Optical
  Access surface ion trap.}, 2016.

\bibitem{GST-improvement-china}
S.~Zhang, Y.~Lu, K.~Zhang, W.~Chen, Y.~Li, J.~N. Zhang, and K.~Kim.
\newblock {Error-mitigated quantum gates exceeding physical fidelities in a
  trapped-ion system}.
\newblock {\em Nature Communications}, 11(1):1--8, 2020.

\bibitem{Song2019}
C.~Song, J.~Cui, H.~Wang, J.~Hao, H.~Feng, and Y.~Li.
\newblock {Quantum computation with universal error mitigation on a
  superconducting quantum processor}.
\newblock {\em Science Advances}, 5(9):eaaw5686, 2019.

\bibitem{google-RB}
J.~Kelly, R.~Barends, B.~Campbell, Y.~Chen, Z.~Chen, B.~Chiaro, A.~Dunsworth,
  A.~G. Fowler, I.~C. Hoi, E.~Jeffrey, et~al.
\newblock {Optimal quantum control using randomized benchmarking}.
\newblock {\em Physical Review Letters}, 112(24):1--5, 2014.

\bibitem{noise-coherence-2015}
J.~Wallman, C.~Granade, R.~Harper, and S.~T. Flammia.
\newblock {Estimating the coherence of noise}.
\newblock {\em New Journal of Physics}, 17(11):1--10, 2015.

\bibitem{unitary-RB}
S.~Sheldon, L.~S. Bishop, E.~Magesan, S.~Filipp, J.~M. Chow, and J.~M.
  Gambetta.
\newblock {Characterizing errors on qubit operations via iterative randomized
  benchmarking}.
\newblock {\em Physical Review A}, 93(1):1--6, 2016.

\bibitem{Endo2018}
S.~Endo, S.~C. Benjamin, and Y.~Li.
\newblock {Practical Quantum Error Mitigation for Near-Future Applications}.
\newblock {\em Physical Review X}, 8(3):31027, 2018.

\bibitem{OBrien2017}
T.~E. O’Brien, B.~Tarasinski, and L.~DiCarlo.
\newblock {Density-matrix simulation of small surface codes under current and
  projected experimental noise}.
\newblock {\em npj Quantum Information}, 3(1):39, 2017.

\bibitem{intro-GST}
D.~Greenbaum.
\newblock Introduction to quantum gate set tomography.
\newblock {\em arXiv:1509.02921v1}, 2015.

\bibitem{open-pulse}
D.~C. McKay, T.~Alexander, L.~Bello, M.~J. Biercuk, L.~Bishop, J.~Chen, J.~M.
  Chow, A.~D. C{\'{o}}rcoles, D.~Egger, S.~Filipp, et~al.
\newblock Qiskit backend specifications for openqasm and openpulse experiments.
\newblock {\em arXiv:1809.03452}, 2018.

\bibitem{inverseCP}
T.~F Jordan.
\newblock {Maps and inverse maps in open quantum dynamics}.
\newblock {\em Annals of Physics}, 325(10):2075--2089, 2010.

\bibitem{Temme2017}
K.~Temme, S.~Bravyi, and J.~M. Gambetta.
\newblock {Error Mitigation for Short-Depth Quantum Circuits}.
\newblock {\em Physical Review Letters}, 119(18):1--5, 2017.

\bibitem{pygsti}
E.~Nielsen, K.~Rudinger, T.~Proctor, A.~Russo, K.~Young, and R.~Blume-Kohout.
\newblock {Probing quantum processor performance with pyGSTi}.
\newblock {\em arXiv:2002.12476}, 2020.

\bibitem{DRB}
T.~J. Proctor, A.~Carignan-Dugas, K.~Rudinger, E.~Nielsen, R.~Blume-Kohout, and
  K.~Young.
\newblock {Direct Randomized Benchmarking for Multiqubit Devices}.
\newblock {\em Phys. Rev. Lett.}, 123(3):30503, 7 2019.

\bibitem{PhysRevLett.119.130502}
T.~Proctor, K.~Rudinger, K.~Young, M.~Sarovar, and R.~Blume-Kohout.
\newblock {What Randomized Benchmarking Actually Measures}.
\newblock {\em Phys. Rev. Lett.}, 119(13):130502, 9 2017.

\end{thebibliography}
\end{document}